\documentclass[conference]{IEEEtran}
\usepackage{cite}
\usepackage{graphicx}
\usepackage[hyphens]{url}
\usepackage{flushend}
\usepackage{url} 
\usepackage{hyperref}

\begin{document}
\title{Longitudinal Analysis of Android Ad Library Permissions}
\author{\IEEEauthorblockN{Theodore Book, Adam Pridgen, Dan S. Wallach}
\IEEEauthorblockA{Rice University\\\{tbook, dso, dwallach\}@rice.edu}}
\maketitle
\begin{abstract}
This paper investigates changes over time in the behavior of Android ad libraries.  Taking a sample of 114,000 apps, we extract and classify their ad libraries.  By considering the release dates of the applications that use a specific ad library version, we estimate the release date for the library, and thus build a chronological map of the permissions used by various ad libraries over time.  By considering install counts, we are able to estimate the number of times that a given library has been installed on users' devices.  We find that the use of most permissions has increased over the last several years, and that more libraries are able to use permissions that pose particular risks to user privacy and security.
\end{abstract}
\section{Introduction}
The Android operating system is one of the primary mobile device platforms worldwide, accounting for 70\% of smartphone shipments, according to industry sources~\cite{strategy-android-market-share}.  It also supports a vibrant advertising industry, with dozens of advertising agencies providing ad libraries installed in hundreds of thousands of free applications.  The diversity of these libraries and the availability of the applications for research makes Android an ideal platform to study the impact of advertising libraries on mobile device security.\par
Similar to the web domain, the dominant monetization model for applications on the Android platform consists of free applications where the developers receive compensation through the sale of ads~\cite{leontiadis2012don}.  The advertising model on which this is built is based on Java libraries supplied by advertising agencies which are bundled into application packages.  These libraries communicate with servers controlled by the advertising agency, sending request information, receiving and displaying an advertisement, and handling user interaction with the ad, including redirects to content and accounting functions.
These libraries execute as a binary portion of the host application.\par
Thus, there are three principal interfaces for an ad library: with the rest of the application, with Internet resources (principally the ad provider's servers), and with the Android operating system.  It is with the third of these interfaces that this paper will be concerned.\par
\par\textit{The Android Permission System:}
The Android operating system limits application privileges through a permission system.  Privileges are requested by the application developer in a manifest which is incorporated into the application package, and approved by the user when the application is installed.  The user may only choose to accept all of the permissions or not install the app --- they can not grant or deny specific permissions.  The Android API reference lists 130 permissions, which regulate the ability of an app to make various system API calls.  The functions covered by a permission range from the clear to the obtuse --- for example, the read phone state permission also gives access to a commonly used device identifier.\par
Because the granularity of these permissions is at the level of the application, an advertising library necessarily shares in all of the privileges of its hosting application.  As an ad library is able to probe for the presence of a particular permission, it is potentially able to make use of permissions in its host app on an opportunistic basis~\cite{stevensinvestigating}.  Of course, an ad library can also require the presence of certain permissions to function.\par
Concerns have been raised about the interest and ability of users to make sound security judgments on the appropriateness of an app's behavior based on the permission system\cite{felt2012android}.  Indeed, the permission system has not prevented significant negative publicity related to Android apps making inappropriate use of personal data, even when the user approved the permissions requested in the manifest~\cite{appsSteal}.
\par\textit{Research Goals:}
A number of studies have documented that Android ad libraries are able to use application permissions to gain access to sensitive user data~\cite{grace2012unsafe}.  However, these studies have focused on the state of the Android ad ecosystem at a single point in time.  This paper seeks to extend that work by looking at change over time in library permission usage.  By doing so, developments in the ad ecosystem driven by technical progress, privacy concerns, and financial incentives become more explicit.\par

\section{Methodology}
\begin{figure}
\centering
\includegraphics[width=\columnwidth]{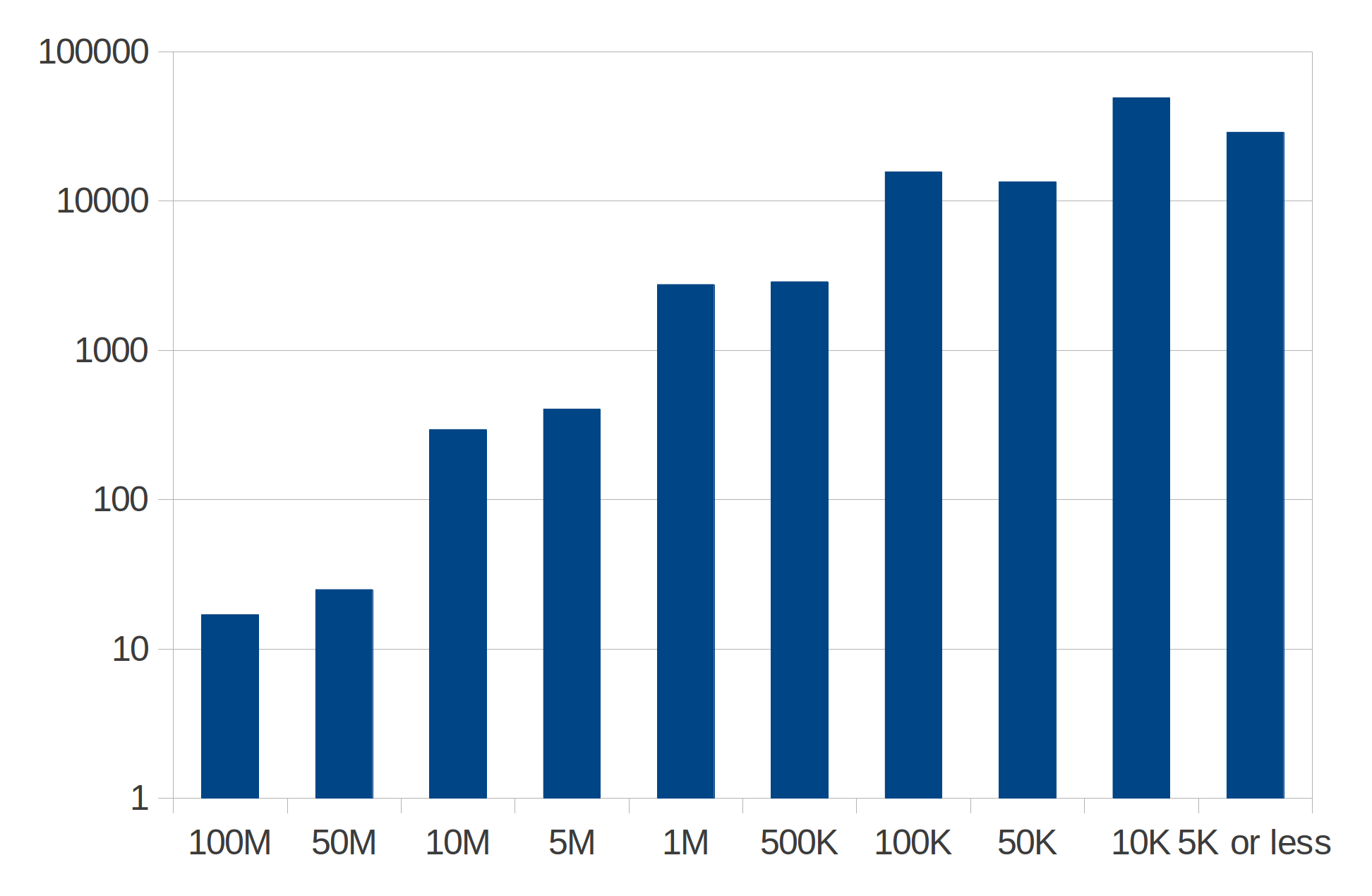}
\caption{Distribution of Sample by Number of Installs}
\label{installs-by-appcount}
\end{figure}
In order to conduct this analysis, we needed to obtain an appropriate sample.  A collection of 114,000 applications were downloaded from Google Play --- the primary Android application store.  All were free applications, which are, presumably, the type most likely to contain advertisements.  We selected the most popular apps from a database of 375,000 applications extracted from Google Play, which contained information on the number of installs and release date for each app.  This database reflects roughly half of the over 675,000 apps that Google reports on the Google Play overview page~\cite{playOverview}.  We assessed popularity by selecting applications on the basis of number of installs, with preference given to higher numbers of installs.  This enabled the sample to reflect the installed app base more accurately.  Figure~\ref{installs-by-appcount} shows the distribution of the sample according to the lower bounds on the total number of installs for each app, as provided by Google Play.\par
After collecting the applications, we disassembled them using Gabor Paller's Dedexer.  We manually identified package names for 68 different ad libraries by parsing through the collection of disassembled apps.\footnote{Included in the libraries were several ad mediation and analytics libraries}  We then automatically identified all ad libraries matching our known package names within the entire sample.  It should be noted that this method necessarily produces an incomplete collection of ad libraries.  It omits any libraries for which we did not have a package name, and those libraries (such as AirPush) that obfuscate their package names for some versions.  We extracted 108,000 ad library instances, an average of approximately one library per app, although some apps contained many libraries, and others contained none.  We then hashed each library to generate a unique identifier corresponding to the library version.  See below for some complications.  In this way, we were able to identify the library name and separate the libraries by version without needing any special knowledge of each library's internal structure.  We then grouped all libraries with the same version hash.\par
\textit{Dating:}
Once we had identified library names and versions, it became possible to date them.  Because we knew the release date for each of the apps, it was safe to assume that the library version was released before the first app which used it.  Thus we were able to establish a \textit{terminus ad quem} or latest possible release date for each library version by assigning it a date equal to the release date of the earliest app that made use of that library.\par
In order to verify the accuracy of this approach, we extracted library version codes from the Google AdMob libraries we identified.  This enabled us to compare the calculated release dates with the actual release dates as reported by Google~\cite{adMobReleaseNotes}.  On average, our calculated date was 18 days before the official release, while several major versions were dated within one day of the official release.  We also identified several AdMob versions not listed by Google in its release history.  The earlier dates could be a result of beta/testing releases of the APKs, or of inaccuracy in the app release dating provided by Google.  For libraries included in fewer applications, our dating approach will necessarily be less precise.  However, the demonstrated accuracy of the AdMob dating suggests that our methodology has value in practice, enabling us to survey effectively a large number of libraries without having to reverse engineer them.\par

\textit{Measuring Permisson Usage:} We measured permission usage by analyzing the disassembled libraries.  A list of system calls requiring permissions was obtained from the API call mappings produced by the PScout tool developed by Au, et al.~\cite{PScout}.  Our method of generating permission mappings is similar to that used by Felt, et al.~\cite{felt2011android} in their work on Android permissions, however, we considered permission usage only within the ad library code, and not within the entire application.  Because we only consider static API calls, our analysis is conservative, and would miss API calls invoked via reflection, direct generation of Dalvik bytecode, dynamically downloaded bytecode, etc.  Using this method, however, we were able to identify a large number of system calls and the permissions required to make those calls.\par
Because we were interested in the libraries, themselves, we did not conduct control-flow analysis to identify whether these library API calls were, in fact, invoked in any particular app.  Nor did we correlate the permissions necessary to use the calls with the permissions present in the manifest for any particular app.  Rather, by showing the presence of API calls requiring a certain permission, we showed that the library was capable (under some set of circumstances) of making use of that permission.  Therefore, for API calls that require one permission out of a set of multiple permissions to function, we included all of the permissions in the set as permissions that the library was capable of using.  While the permissions used by some API calls, such as some of the location calls, vary depending on the input given to the call, most API calls require a fixed set of permissions that do not change depending on their input.  In cases where the input to the API call influences which permissions are checked, we include all permissions that might be used, admittedly a maximal set.  While it would be interesting to see which applications actually contain control flow paths that reach these API calls, we chose to focus on the structure of the libraries, rather then the way they were employed in individual applications.

\par\textit{Identifying Unique Ad Library Releases:}
We encountered some difficulties in identifying ad library releases by hashing the decompiled code.  Due to the way that an Android APK is assembled, virtual register numbers within each library instance varied, based on the other components of the app.  We compensated for this by ignoring the actual names of the virtual registers and only considering the presence of a field.\par
Beyond this, however, additional challenges presented themselves.  Because the process of assembling an Android application, especially when obfuscating and optimizing software such as ProGuard is used, may omit or rename portions of the library, a single library version may vary in size and content after being included in an app.  This means that a single library release may produce multiple installed versions, each with different capabilities and signatures.  Rather than attempting to correlate which version hashes derived from the same library release, we allowed different hashes to stand independently.  While this made it impossible for us to count accurately the total number of releases of a particular ad library, it allowed us to measure permission use and date library releases as desired.\par
Some negative impacts from this methodology should be noted.  Some of the code of the original ad library release may not be included in a given app, including API calls.  This means that it is possible for our methodology to under-measure permission usage of a given ad library version release.  Additionally, because a library release that becomes fragmented into multiple versions will have fewer apps recorded for each version, the dating becomes less accurate.  However, to the extent that our dating is correct, all of the different hashes produced by a single release should be dated to approximately the same period.  Because our analysis takes the union of all permissions used by library versions dated to a given month, the permissions available to the most complete version of the library are the ones actually measured.

\section{Analysis}
Having identified library names and versions, dated them, and mapped their permission usage, it then became possible to map the change in permission usage by various ad libraries over time.  We derived the following results from the data.

\subsection{Permission Usage by Library}
In order to develop a preliminary understanding of the data, it was useful to consider the number of permissions used by each library.  While for many libraries, the number of permissions changed over time, the maximum number of permissions (or sets of equivalent permissions) that we found any library able to use was 15, exploitable by some versions of MobClix.\footnote{Those permissions included \textsc{internet, access network state, wake lock, get tasks, access fine location, access coarse location, camera, read phone state, vibrate, write social stream, write contacts, read social stream, read contacts, read sync settings,} and \textsc{get accounts}.  Note that multiple sets of permissions can give access to a single API call; for example, the various \textsc{social stream} and \textsc{contacts} permissions each allow a library to read a user's contacts.}  The average library was able to make use of 3.3 permissions.  The minimum number of permissions that a library could exploit was one, with the \textsc{internet} permission being required by all libraries.  While some libraries are much more popular than others, previous researchers have generally not calculated the number of installs for each library.  We include this per-library data to enable better comparison of our results with theirs.

\begin{figure}[h]
\centering
\includegraphics[width=\columnwidth]{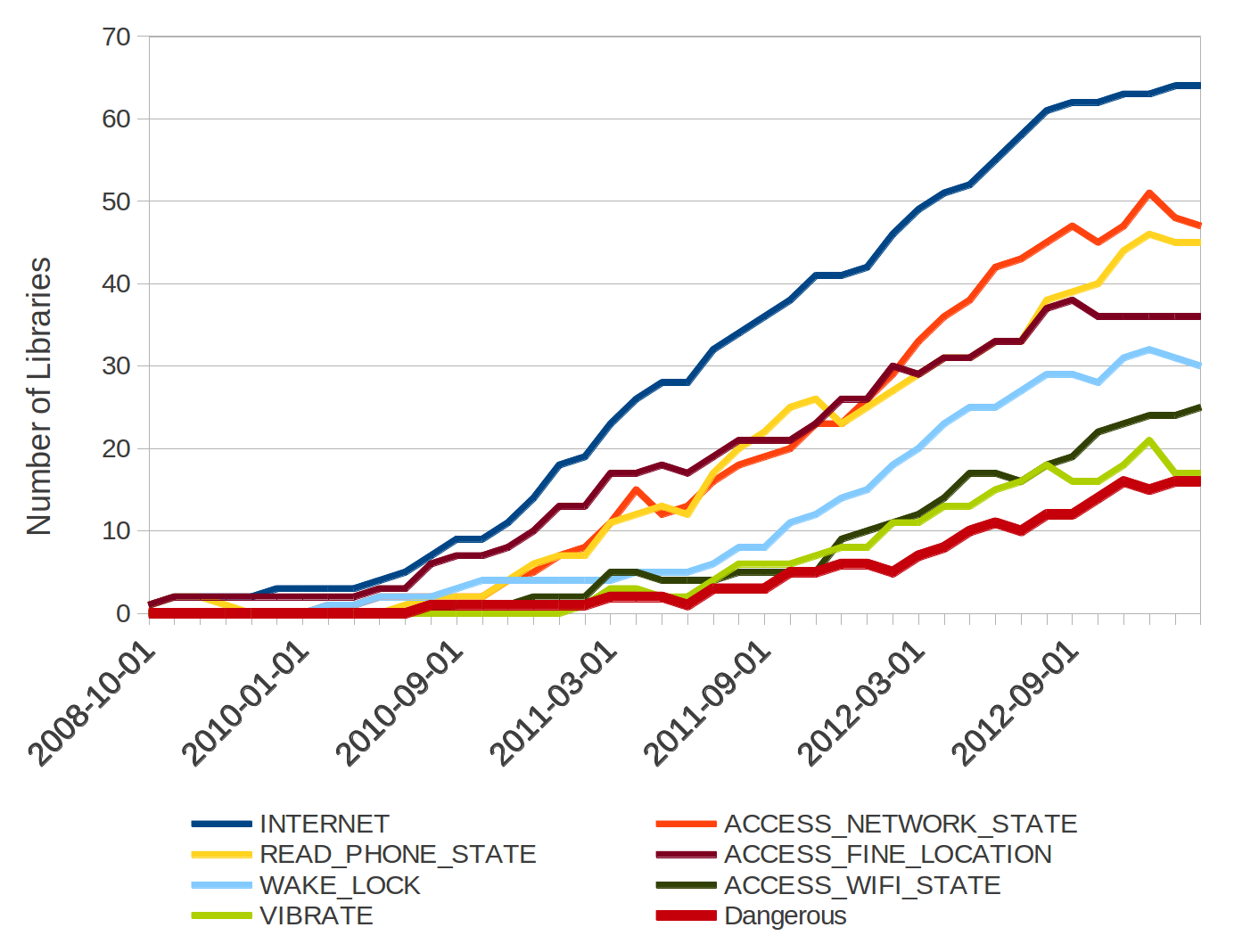}
\caption{Number of Libraries Able to Use Permission}
\label{LibraryCountByPermission}
\end{figure}
\subsection{Permission Usage over Time}
More interesting than the static data is the changing number of permissions that a single library can utilize.  By simply examining the number of libraries able to use a single permission, we are able to see which permissions are accessible by the greatest number of libraries.  Figure~\ref{LibraryCountByPermission} shows the number of libraries capable of using the seven most common permissions.\footnote{In the cases where multiple versions of a library dated to a single month, we took the union of all the permissions called for by any version of the library.}  A number of less common, but potentially dangerous permissions are grouped together under the heading of ``Dangerous".  A library that used one or more of these permissions was categorized as using dangerous permissions.\par
We categorize permissions as potentially dangerous when they provide access to personal user information, make it possible to directly monitor the user's activity, allow the library to make operations that cost money, or provide access to the system state.  For more information, see section~\ref{dangerous}.\par
It is important to note that the ability of an ad library to utilize these permissions does not necessarily mean that they were being abused.  With appropriate user confirmation, many of them could have legitimate uses.  However, as these permissions also enable various sorts of abuse, they are worthy of special scrutiny.\par
\begin{figure}
\centering
\includegraphics[width=\columnwidth]{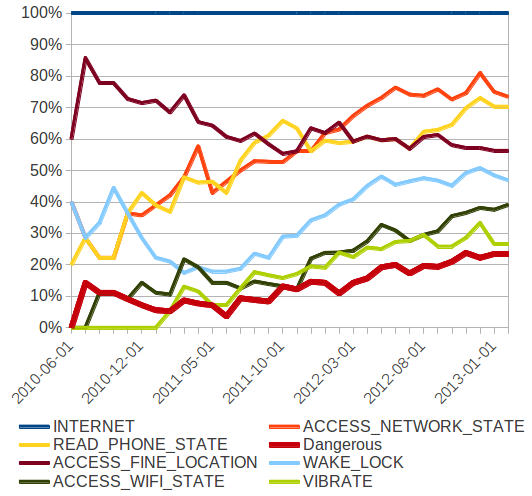}
\caption{Percent of Libraries Able to Use Permission}
\label{LibraryPercentByPermission}
\end{figure}
While Figure~\ref{LibraryCountByPermission} gives a count of the number of libraries using a particular permission, it may be more useful to present this data on a percentage basis.  Figure~\ref{LibraryPercentByPermission} represents the percent of known libraries using a given permission for the months where we had data on five or more libraries.  The data becomes more precise for later months where we were able to measure additional ad libraries.  As can be seen, most permissions, including the dangerous permissions, show an increasing trend, a potential cause for concern.\par
Although we did not conduct a comprehensive analysis of all API calls requiring permissions, some notes may be made on the most popular permissions.  The universal use of the \textsc{internet } permission corresponds to the ad library's most basic function --- downloading and displaying advertisements.  The \textsc{access network state} permission is used by a number of libraries to call \textsf{\small{android.net.ConnectivityManager.\allowbreak getActiveNetworkInfo}}, presumably in order to determine the connection over which they are requesting data.  The \textsc{read phone state} permission is necessary for retrieving a device ID from the telephony manager.  Likewise, \textsc{access wifi state} may be used to obtain the MAC address of the device to help identify it (and its user) uniquely.  \textsc{wake lock} is used by some video playback API calls.    \textsc{vibrate} is self-explanatory.  While \textsc{access fine location} is also self-explanatory, it is worth noting that all library calls surveyed that made use of system calls for location data used calls that could use either \textsc{access fine location} or \textsc{access course location}, depending on which permission was available.\par

\subsection{Library Market Share}
Of course, data that only tracks the number of libraries using a certain permission is somewhat artificial, as some libraries are much more popular than others.  In order to understand better the permissions accessible by the ad libraries that might be present on a randomly chosen device, we corrected the data based on the popularity of a given ad library.\par
\begin{figure}
\centering
\includegraphics[width=\columnwidth]{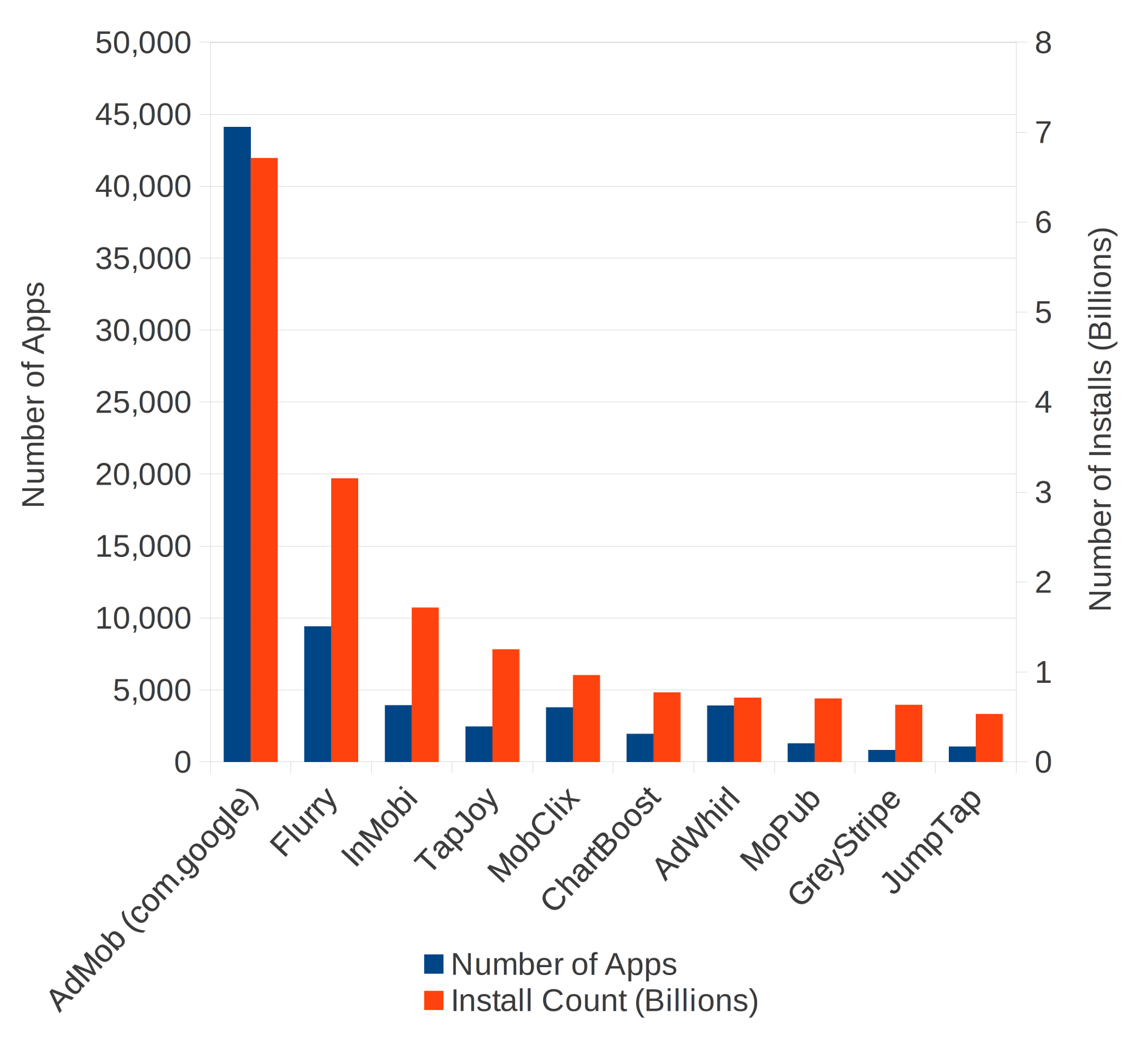}
\caption{Top 10 Libraries by Installs.  The number of apps containing a given library is presented on the primary Y axis, and the number of installs for a given library on the secondary Y axis.}
\label{popular}
\end{figure}

Library popularity can be measured in two different ways --- by the number of apps using a library, and by the cumulative number of installs for a given library across all apps that include it.  Figure~\ref{popular} shows the number of apps within our dataset where each library was detected, and an estimate of the number of installs for each library.  The top 10 libraries represent 71\% of all measured installs.\par
Google reports app install data based on ranges with a factor of two or five between the top and the bottom of the range. (e.g., 1,000--5,000 installs or 5,000--10,000 installs.)  For our analysis, we chose the low end of the range as a conservative estimate of the number of installs for a given app.  Because we do not have more precise install counts, the actual numbers may be greater by as much as a factor of five.  Additionally, because the install count only considers apps within our dataset, when other apps are included, the numbers undoubtedly become larger.  Nevertheless, these counts give us an effective means to compare the relative popularity of different libraries.\par
Because our set of apps focused on those apps with higher numbers of installs, this data should accurately reflect the popularity of ad libraries among the most popular apps.  It is possible that a different distribution is found among less popular apps.  However, as the more popular apps are, by definition, more likely to be found on any given device, we believe that this distribution is reflective of the distribution of ad libraries likely to be found on a user's device.\par
It is worth noting that while a few libraries (particularly Google's AdMob) have a clearly dominant position in the marketplace, there are a large number of libraries with significant market share.  The top 25 libraries have approximately 90\% market share, while the top 33 have 95\% market share.  It is, of course, likely that we failed to identify some ad libraries, although we assume that we were able to correctly identify the most popular ones.\par
Given the fact that there are dozens of different ad libraries with significant market share (nearly all of the libraries we surveyed had tens if not hundreds of millions of installs), it becomes clear that the security of a user's data and device depends on the behavior of a large number of principals.  While we did not attempt to estimate the number of libraries likely to be found on a typical device, this data shows that there is certainly a possibility for many libraries to be present, each of which presents security concerns.

\subsection{Install Weighted Permission Usage}
\begin{figure}
\centering
\includegraphics[width=\columnwidth]{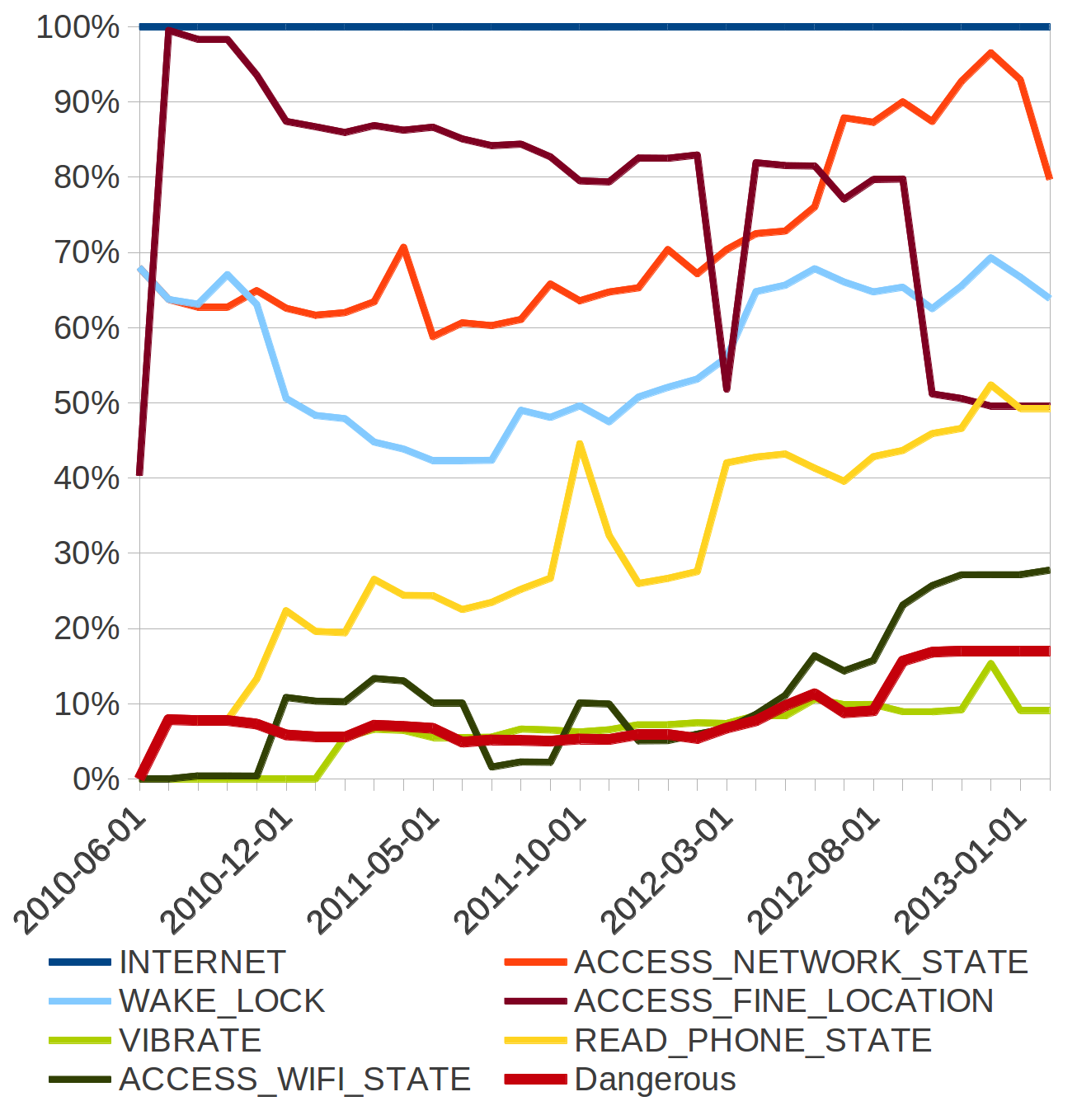}
\caption{Percent of Library Installs Able to Use Permission}
\label{permissions-by-installs}
\includegraphics[width=\columnwidth]{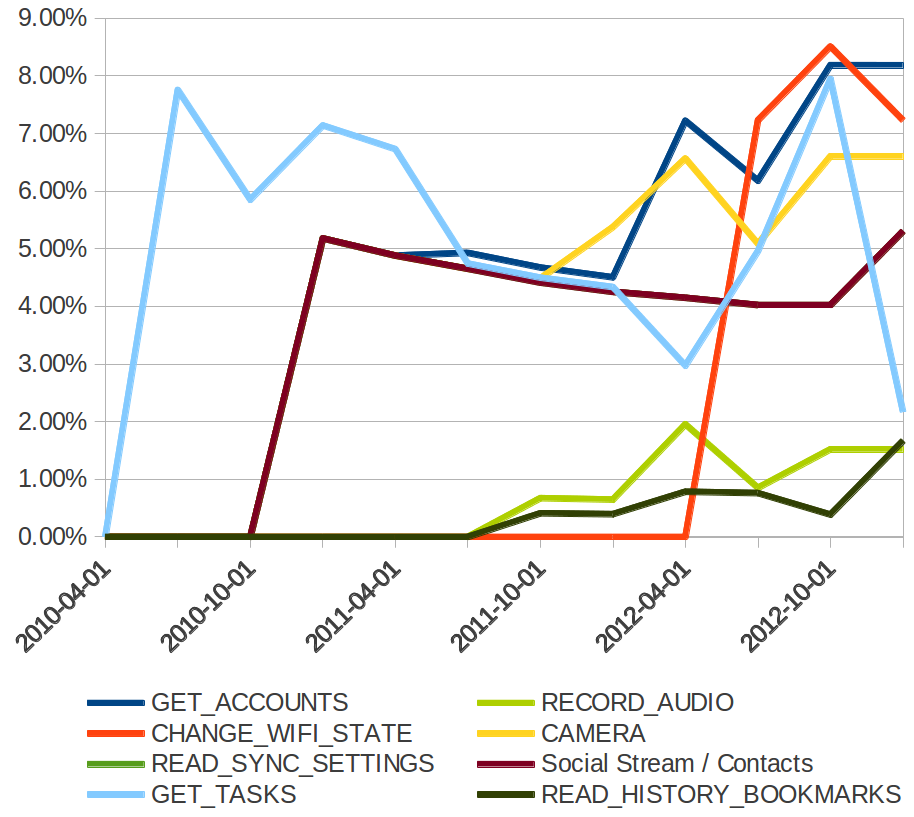}
\caption{Dangerous Permissions: Percent of Library Installs Able to Use Permission}
\label{dangerous-by-installs}
\end{figure}

Using library popularity data to weight the various libraries, it then becomes possible to see the permissions that might be accessible to a ``typical" ad library install.  Because the granularity of the data made it difficult to reconstruct the historic market share of different ad libraries, we made the assumption that library market share remained constant over the period studied.  The results of this analysis is shown in Figure~\ref{permissions-by-installs}.\footnote{The downward spike in the \textsc{access fine location} permission in March of 2012 is due to version 5.0.0 of AdMob.  We did not detect any location code in the few samples of that version that we collected, although we did find it in version 6.01, which was dated slightly later.}  We observe that the general trends are similar to the trends seen when we considered individual libraries, but the absolute values for some permissions are different.  In particular, \textsc{access wifi state} is used in a lower percentage of installs, as is \textsc{vibrate}.  \textsc{access fine location}, however, is more common, mainly due to the use of \textsf{\small{android.location.LocationManager.\allowbreak getLastKnownLocation}} in some versions of AdMob.  The removal of that API call accounts for the significant drop in that permission in the middle of 2012.  It can be seen that the use of dangerous permissions is somewhat less common, as they are not used by the largest ad libraries.  Likewise, the permissions used to generate unique tracking IDs (\textsc{read phone state} and \textsc{access wifi state}) are somewhat less common.\footnote{Android supplies an install specific ID that can be accessed without any permissions (\textsf{\scriptsize{Settings.Secure.ANDROID\_ID}}). However, this ID may change upon factory reset, or on a rooted phone.  Occasional implementation bugs may result in it being missing, or the same across multiple devices.  These shortcomings may encourage library designers to seek other IDs.}  The \textsc{vibrate} permission is much less commonly used.

\subsection{Dangerous Permissions}
\label{dangerous}
The subset of permissions which we labeled ``Dangerous Permissions" comprises a relatively small set of the permissions used by Android ad libraries, but is still of particular interest.  Figure~\ref{dangerous-by-installs} shows the install weighted prevalence of the various dangerous permissions among the libraries studied.  Although the data is noisy due to the relatively small number of libraries using these permissions, a generally increasing trend can be seen. A brief review of these permissions gives some indication of their use.  \textsc{get tasks} refers to processes, not a user to-do list, and seems often to be used for API calls such as \textsf{\small{android.app.ActivityManager.\allowbreak getRunningTasks}} and \textsf{\small{android.app.\allowbreak Activity\allowbreak Manager.\allowbreak getRecentTasks}}.  These may be used in conjunction with \textsf{\small{PackageManager}}, which will return the list of installed apps without requiring any permissions, in order to determine which apps are installed on the device, and which are regularly used.  This data represents a potential privacy breach, and could represent a security weakness if the apps are scanned to identify apps that have potential security weaknesses.\par
The \textsc{social stream, contacts}, and \textsc{sync settings} permissions are used by calls such as \textsf{\small{android.provider.ContactsContract.\allowbreak Contacts.\allowbreak getLookupUri}} to obtain information about the user's contacts.  The \textsc{get accounts} permission allows the library to identify if a user has an account on their device with a provider such as Google or Facebook through an API call such as \textsf{\small{android.accounts.AccountManager.\allowbreak getAccountsByType}}.  While full credentials for a user's accounts are not returned, the account name is.  For example, on Google accounts, the account name is the user's email address, allowing the ad agency to potentially send e-mail to the user, while providing another unique identifier.  \textsc{change wifi state} is used by some ad libraries to scan for WiFi access points, which has potential value in determining the user's location.  The use of the \textsc{camera} permission allows the library to control the camera directly, without the need to send an intent to the camera application.  Some libraries advertise the ability for the user to interact with the ad by taking pictures and uploading them to the ad server.  The \textsc{record audio} permission is likewise used by some ad providers to record audio.  Likewise, the \textsc{read history bookmarks} is used to read the web browser bookmarks through the \textsf{\small{android.provider.Browser.getAllBookmarks}} interface.  The privacy issues involved in this are obvious.\footnote{In addition to the permissions included in the graph, several other, extremely rare, permissions were included in the set of dangerous permissions.  They are: \textsc{kill background processes, send sms, write history bookmarks, write external storage, manage accounts, write sync settings, authenticate accounts, restart packages, bluetooth, bluetooth admin, broadcast sticky}, and \textsc{modify audio settings}.}
\subsection{Comparison With Other Datasets}
In order to further assess the accuracy of our results, we compared them with a library of 5,000 Android apps retrieved from the Android Market (now Google Play) in May of 2011, as part of the AdSplit project~\cite{shekhar2012adsplit}.  When analyzed through the same process, these apps yielded 3,374 ad libraries, a slightly smaller ratio than the principal dataset.  When the permissions used by those libraries were analyzed, they were substantially the same as the inferred values for May of 2011.  The greatest variation was in the \textsc{vibrate} permission, which showed up in 15\% of libraries in the 2011 sample, while the 2013 data implied a 7\% rate.  Of course, the 2011 sample undoubtedly contains many apps older than the sample date.  Unfortunately, however, metadata to track the app release dates was not collected with the AdSplit sample, so we can not attempt to reconstruct the release dates for the ad libraries found therein.

\subsection{Overall Trends}
Whether considered on a per-library or per-install basis, it is clear that the number of permissions which the average ad library is capable of using has been steadily increasing over the two and a half year period for which we had reasonable data.  Particular growth is seen in the \textsc{read phone state} and \textsc{access network state} permissions, although almost all permissions show an upward trend.  Presumably, this is because ad library developers find that their libraries are more effective when they make use of these additional permissions.\par
Unfortunately for users, however, most of the permission creep seems directed not at making the ad libraries work more smoothly and efficiently (with the possible exception of the \textsc{access network state} permission), but at extracting additional data about the user.  Permission growth occurs largely in the permissions needed to uniquely identify the device (and, by extension, the user),  such as \textsc{read phone state} and \textsc{access wifi state}, and in permissions that make ads more obtrusive, such as \textsc{vibrate}.  It is possible that some of the permission growth relates to an increasing ability of ad libraries to incorporate rich media such as video --- in particular, the \textsc{wake lock} permission seems to be related to video API calls.  However, most seems to be simply related to the collection of additional data on the user.\par
Within the area of permission growth the permissions identified as dangerous represent a significant area of concern.  While the libraries that make use of these permissions comprise a small portion of the total sample, a growing number of libraries do make use of them.  The same dynamics seem to be at work as with the more common permissions --- a desire for a more intrusive ad experience coupled with a desire to obtain more information regarding the user.  MobClix, one of the largest users of permissions, seeks to attract advertisers by advertising the ability to access contacts, the user's media library, GPS, calendar, camera, e-mail, and SMS~\cite{mobclixRichMedia}.\par
The principal exception to increasing permission use is a steady decline in the number of libraries making use of the \textsc{access fine location} and \textsc{access course location} permissions.  Google has played a significant role by removing location code from its AdMob library (although app developers can still choose to pass location information manually) but this change has also been reflected in a large number of other libraries.  It may be that ad providers are able to obtain sufficiently detailed location information from the phone's IP address, or that the commercial value of location information is less than was once expected, but it is also possible that users' privacy concerns have created an environment where app developers prefer to use ad libraries that do not include location code.  Overall, however, ad library permission usage remains on an increasing trend.

\section{Related Work}
A number of authors have analyzed Android applications to assess security vulnerabilities.  For example, Enck et al.~\cite{enck2011study} conduct a broad study of security vulnerabilities in 1,100 disassembled apps.  Gibler et al.~\cite{gibler2012androidleaks} use static analysis to identify potential leaks of personal information in Android applications.\par
A significant amount of research has been conducted into Android ad libraries.  Stevens, et al.~\cite{stevensinvestigating} analyze 13 ad libraries, comparing their permission usage with their documented permission requirements, and show privacy weaknesses related to user tracking as well as system APIs exposed to Javascript exploits.  Grace, et al.~\cite{grace2012unsafe} conduct a similar analysis on 100 ad libraries selected from 100,000 apps.  They likewise identify permission usage and other security risks, such as the potential to load and execute arbitrary bytecode through the ad interface.  Neither of these papers, however, consider the development of ad libraries over time, or even make reference to variations among different versions of the same library.\par
Additional research has targeted other aspects of the Android ad ecosystem.  The data traffic\cite{zhang2012expensive} and energy\cite{pathak2012energy} usage involved in displaying ads has been examined.  Leontiadis, et al.\cite{leontiadis2012don} look at the other side of the issue, considering the importance of advertising, including access to personal information, in maintaining the Android app ecosystem, while examining various ways of regulating and monitoring an Android application's access to personal data.  Vallina Rodriguez et al.\cite{vallina2012breaking} study mobile ads through an analysis of network traffic on a major European mobile carrier, considering energy implications and ways of optimizing the ad delivery system.\par
Significant work has also been done on resolving the security issues involved in the current Android ad model.  One of the most prominent approaches has been to advocate separating ads from applications, enabling isolation of permissions and data~\cite{shekhar2012adsplit}\cite{pearce2012addroid}.  In particular, the AdSplit framework considers the value of providing advertisers with protection against fraudulent behavior on the part of app developers, while at the same time separating apps and ad libraries, allowing each to function with separate permissions.

\section{App Store Policing}
After we collected our original data, Google conducted a purge of some 60,000 apps from Google Play\cite{playPurge}.  We were interested in seeing how the apps in our sample fared, and so we re-surveyed the apps to see which had been removed.  While many of the apps re-appeared on Google Play within a matter of days (presumably with changes made to address Google's concerns), we were able to identify 9,980 apps from our sample (almost 10\%) that were no longer available.  While some of the apps may have been removed by their publisher, it seems likely that many were removed by Google.\par
After identifying the missing applications, we went back to our original data set to see what ad libraries were used in those apps.  We found that apps containing certain libraries were much more likely to be removed than others.  Some of the libraries whose apps were disproportionately removed are included in Figure ~\ref{purged}.    The column ``Missing" indicates the number of apps containing a library that were removed.  ``Original" gives the number of apps in the original sample that contained the library.  The next column gives the percentage of the total number of apps that were removed.  ``Permissions" gives the total number of permissions that we identified as being usable by the library.\par
While we made this discovery too close to publication time for more in--depth analysis, further research could determine if the restored versions of these apps lacked these libraries, suggesting that removal of these apps is related to the behavior of their ad libraries.  While we have no inside knowledge of what caused Google to select some of these apps for being removed, it's possible that poorly-behaved advertising libraries might be complicit in their host applications' deletion. By looking at the ``permissions" column (fig.~\ref{purged}), we can see that these advertising libraries seem to use an unusually high number of permissions. However, the fact that some instances of apps using these libraries were deleted while others remain suggests that Google isn't simply banning specific advertising libraries.  \par
Many of the libraries included in Figure ~\ref{purged} exhibited additional negative behavior: EverBadge and AirPush place launch icons on the user's home screen; AirPush, MobPartner, and SendDroid all publicize the ability to send advertisements to users in the form of push notifications.  Ongoing analysis will be able to determine the shifting nature of ad libraries, as well as Google's response to the challenges posed by them.
\begin{figure}
\begin{tabular}{l | r | r | r | r }
& Missing & Original & Removed & Permissions\\
\hline
EverBadge  & 81 & 134 & 60.5\%	&2\\
Hunt Mobile& 161 & 354 & 45.5\%	&3\\
AirPush \dag & 801 & 1966 & 40.7\%	&9\\
SendDroid & 417 & 1,335 & 31.2\%	&8\\
Waps  & 57 & 192 & 29.7\%	&8\\
Tapit & 130 & 458 & 28.4\%	&7\\
AdsMogo  & 19 & 67 & 28.4\%	&8\\
Adfonic  & 186 & 767 & 24.3\%	&5\\
RevMob  & 166 & 706 & 23.5\%	&6\\
\hline
Average & & & 11.6\% & 5.62
\end{tabular}
\caption{Libraries found in apps removed from Google Play.}
\label{purged}
\end{figure}
\section{Future Work}
Much remains to be done in tracking the development of the Android ad ecosystem.  Ongoing sampling of Google Play would allow the construction of a dataset of apps providing greater detail for diachronic studies.  Similar sampling of third party app stores would provide an additional longitudinal dataset of significant value.\par
Additionally, more precise measurement of the available data may yield interesting insights.  Static analysis could reveal the conditions under which the various API calls are made, and help to determine if they are explicitly triggered by the user, an important measurement for privacy questions.  It is conceivable that, while apps are capable of making more sensitive API calls, they are also making them in a more responsible fashion.\par
Furthermore, a more thorough understanding of the information exchanged between the app and its libraries, and between the libraries and the ad servers are also needed.  We will seek to address many of these issues in our ongoing research.
\section{Conclusion}
The Android environment was carefully designed to provide security by separating applications from each other and limiting their permissions.  It also provides strong infrastructure for patching and updating applications, either in response to security flaws or to fix bugs and provide new features.  However, the development of ad libraries exposes a significant weakness in that design.  Because an ad library is bundled within an application, it operates with the same permissions as the application.  There is also no way for the library developer to push updates, aside from releasing a new library version and encouraging developers to adopt it.  It was this flaw that made it possible for us to sample older library versions by looking at apps currently released for download.\par
Our research showed that ad libraries are increasingly taking advantage of permissions that may be requested by the host application.  With the exception of location data, we observed a steady increase in the number of permissions ad libraries were capable of utilizing. Particularly disturbing is the growth in the usage of various dangerous permissions that pose particular privacy risks.  While these permissions are used by only a small number of ad libraries, their use is steadily growing, and should be considered with particular scrutiny.  Given the increasing intrusiveness of advertising libraries on the Android platform, the importance of community norms to protect user privacy becomes ever clearer.  Without an effective response from Google, via its Play Store, the only viable alternative to protect users' privacy would seem to be government regulation.

\section{Acknowledgements}
We thank Adrienne Felt and Daniel Sandler for their thoughtful suggestions, as well as the useful feedback from the anonymous referees. This work was supported, in part, by NSF grants CNS-1117943 and CNS-0964566.

\bibliographystyle{IEEEtran}
\bibliography{most-android}
\section {Appendix 1}
The following chart summarizes the raw data collected for this research.  The first column gives the name of the ad agency, the second column gives the number of apps in which any version of that agency's libraries were found.  The third column gives a lower bound for the number of times that the libraries are installed.  The fourth column gives the total number of permissions used in at least 2\% of the libraries from that agency.  The package name is the package name by which we identified the library in question.
\begin{figure*}
\small
\begin{tabular}{l | l | r | r | r | l }
& Library Name & Number of Apps & Install Count & Number of Permissions & Package Name \\ \hline
1 & AdMob (com.google) & 44,107 & 6,711,025,298 & 5 & com.google.ads \\
2 & Flurry & 9,411 & 3,150,987,621 & 5 & com.flurry \\
3 & InMobi & 3,933 & 1,713,911,340 & 7 & com.inmobi \\
4 & TapJoy & 2,456 & 1,250,599,000 & 5 & com.tapjoy \\
5 & MobClix & 3,782 & 963,004,760 & 15 & com.mobclix \\
6 & ChartBoost & 1,935 & 772,087,100 & 4 & com.chartboost \\
\hline
7 & AdWhirl & 3,913 & 712,401,775 & 4 & com.adwhirl \\
8 & MoPub & 1,283 & 703,792,000 & 5 & com.mopub \\
9 & GreyStripe & 816 & 633,353,550 & 4 & com.greystripe \\
10 & JumpTap & 1,054 & 530,924,930 & 6 & com.jumptap \\
11 & Google Analytics & 1,258 & 530,527,776 & 2 & com.google.analytics \\
12 & AdMob (com.admob) & 6,931 & 507,426,872 & 5 & com.admob \\
\hline
13 & Burstly & 105 & 433,160,050 & 7 & com.burstly \\
14 & SponsorPay & 454 & 393,370,000 & 3 & com.sponsorpay \\
15 & Cauly & 1,504 & 367,011,400 & 7 & com.cauly \\
16 & MobFox & 1,931 & 356,714,780 & 6 & com.mobfox \\
17 & Vpon.com & 687 & 315,392,200 & 6 & com.vpon \\
18 & AppBrain & 2,081 & 311,924,110 & 2 & com.appbrain \\
\hline
19 & Daum.net & 1,888 & 310,528,010 & 11 & net.daum \\
20 & AdMarvel & 611 & 244,899,550 & 6 & com.admarvel \\
21 & AppLovin & 987 & 244,547,600 & 9 & com.applovin \\
22 & Adfonic & 767 & 225,023,320 & 5 & com.adfonic \\
23 & MdotM & 637 & 218,410,710 & 5 & com.mdotm \\
24 & GetJar & 255 & 191,135,000 & 6 & com.getjar \\
\hline
25 & NexAge & 301 & 179,990,000 & 7 & com.nexage \\
26 & InnerActive & 884 & 176,181,360 & 6 & com.inneractive \\
27 & Pontiflex & 520 & 164,456,100 & 6 & com.pontiflex \\
28 & ZestAdz & 690 & 148,549,000 & 1 & com.zestadz \\
29 & MadHouse & 438 & 146,908,110 & 9 & com.madhouse \\
30 & Smaato & 816 & 145,135,170 & 6 & com.smaato \\
\hline
31 & YouMi & 534 & 145,130,600 & 8 & net.youmi \\
32 & mAdvertise.de & 645 & 134,302,100 & 5 & de.madvertise \\
33 & DoMob.cn & 664 & 126,351,770 & 8 & cn.domob \\
34 & Jirbo/AdColony & 362 & 125,025,000 & 4 & com.jirbo.adcolony \\
35 & RevMob & 706 & 106,903,620 & 6 & com.revmob \\
36 & SendDroid & 1,335 & 92,244,650 & 8 & com.senddroid \\
\hline
37 & AirPush\dag & 1,966 & 91,891,250 & 9 & com.airpush \\
38 & Tapit & 458 & 87,377,220 & 7 & com.tapit \\
39 & Medialets & 243 & 81,765,050 & 6 & com.medialets \\
40 & Mediba & 964 & 76,796,010 & 4 & mediba.ad \\
41 & Papaya & 377 & 71,573,750 & 15 & com.papaya \\
42 & Hunt Mobile Ads & 354 & 57,602,320 & 3 & com.huntmads \\
\hline
43 & RhythmNewMedia & 194 & 50,345,050 & 6 & com.rhythmnewmedia \\
44 & TapForTap & 424 & 38,493,210 & 4 & com.tapfortap \\
45 & AdKnowledge & 578 & 34,113,760 & 3 & com.adknowledge \\
46 & Metaps & 134 & 29,850,000 & 4 & net.metaps \\
47 & WiYun & 361 & 21,809,100 & 12 & com.wiyun \\
48 & Vdopia.com & 179 & 21,587,000 & 3 & com.vdopia \\
\hline
49 & Waps & 192 & 21,271,000 & 8 & com.waps \\
50 & AdWo & 336 & 14,353,600 & 4 & com.adwo \\
51 & MoboSquare & 8 & 14,210,000 & 3 & com.mobosquare \\
52 & Vserv & 161 & 11,722,100 & 6 & mobi.vserv \\
53 & WooBoo & 203 & 10,542,100 & 7 & com.wooboo \\
54 & EverBadge & 134 & 10,051,010 & 2 & com.everbadge \\
\hline
55 & AirAd & 138 & 8,856,000 & 5 & com.mt.airad \\
56 & Noqoush/AdFalcon & 102 & 7,361,000 & 5 & com.noqoush.adfalcon \\
57 & Moolah & 268 & 6,190,000 & 4 & com.moolah \\
58 & Kuguo & 160 & 5,354,010 & 11 & com.kuguo \\
59 & BuzzCity & 22 & 5,180,000 & 2 & buzzcity \\
60 & AdsMogo & 67 & 4,890,000 & 8 & com.adsmogo \\
\hline
61 & SellaRing & 94 & 3,331,000 & 7 & com.sellaring \\
62 & StartApp & 22 & 1,535,000 & 3 & com.startapp \\
63 & AdModa & 11 & 1,450,000 & 1 & com.admoda \\
64 & MobPartner & 5 & 1,040,000 & 2 & com.mobpartner \\
65 & QuClix & 9 & 345,000 & 2 & com.quclix \\
66 & lDevelop & 7 & 176,000 & 1 & com.ldevelop \\ \hline  
& TOTALS & 108,852 & 24,274,397,772 \\
\end{tabular}
\end{figure*}
\par \dag Only versions of AirPush that did not have obscured package names were recorded, providing a lower bound on the number of installs.
\end{document}